\documentclass[universe,article,submit,moreauthors]{mdpi}

\firstpage{1} 
\pubvolume{1} 
\issuenum{1}
\articlenumber{0} 
\pubyear{2025} \copyrightyear{2025}

\hreflink{https://doi.org/10.3390/universe0000000}
\doinum{10.3390/universe0000000}
\datereceived{}
\dateaccepted{}
\datepublished{}
\AuthorNames{Amirnezam Amiri}

\Title{Dyson spheres on H--R diagram}

\TitleCitation{Dyson spheres on H--R diagram}

\Author{Amirnezam Amiri $^{1}$\orcidA{}} 
\isAPAStyle{%
\AuthorCitation{Amiri, A.} }
{
\AuthorCitation{Amirnezam Amiri} }{ \AuthorCitation{Amiri, A.} } 

\address{ $^{1}$ \quad Department of Physics, University of Arkansas, 226 Physics Building, 825 West Dickson Street, Fayetteville, AR 72701, USA; amirnezamamiri@gmail.com}

\abstract{The construction of Dyson spheres, megastructures designed to capture the total radiative output of stars, can be one of the most compelling techno-signature scenarios for advanced extraterrestrial civilizations. By considering equilibrium temperatures, we investigate the luminosities and fluxes of Dyson spheres built around two promising classes of host stars: white dwarfs and red M-dwarfs. Using radiative balance arguments and representative stellar parameters, we compute the temperature--radius relationship for full energy interception and place these hypothetical structures on the Hertzsprung--Russell (H--R) diagram to assess their observational signatures. Our results show that Dyson spheres around white dwarfs produce cooler and fainter blackbody emissions, peaking in the near- to mid-infrared, while those around M-dwarfs radiate more strongly but at longer wavelengths. In both cases, the equilibrium temperature decreases as \(R_{\rm D}^{-1/2}\), while the total luminosity and observed bolometric flux remain fixed by the stellar output. These findings highlight the astrophysical suitability of low-luminosity stars as Dyson sphere hosts and provide practical constraints for future techno-signature searches using infrared surveys.}


\begin{document}

\section{Introduction}

The concept of a Dyson sphere, introduced by Freeman Dyson (1960)\cite{Dyson1960}, conceptualizes a megastructure constructed around a star to capture a substantial fraction of its stellar radiation. Initially suggested as a means to identify sophisticated extraterrestrial civilizations by their released infrared heat, Dyson spheres have evolved into a crucial theoretical paradigm in astroengineering and the pursuit of techno-signatures \cite{Wright2014}. These structures absorb stellar radiation and re-emit energy thermally at longer wavelengths, with the effective temperature principally dictated by the sphere's radius and the luminosity of the host star \cite{Dyson1960,Kardashev1964}.

The temperature of a Dyson sphere is crucial for understanding its thermal emission spectrum and observational signatures. According to the the inverse square law, the effective temperature drops as the radius of the sphere increases, assuming a constant stellar luminosity \cite{Carrigan2009,Wright2014}. More substantial spheres release at lower temperatures, shifting emission towards the mid- or far-infrared, hence influencing detectability. Recent infrared sky surveys, such as those by WISE and Spitzer, have utilized this correlation to investigate unexpected infrared excesses that may suggest the presence of megastructures \cite{Wright2014,Griffith2015}.

White dwarfs and red M-dwarfs are particularly promising host stars for Dyson spheres due to their distinct astrophysical characteristics. White dwarfs possess low luminosities and small radii ($\approx 0.01\,R_{\odot}$), which implies that Dyson spheres built around them would be more compact and cooler compared to those around main-sequence stars \cite{Lacki2016,Zuckerman2014}. Red M-dwarfs, the most numerous stars in the Galaxy, exhibit low effective temperatures ($\approx$ 2400--3700 K) and radii ($\approx 0.1-0.6\,R_{\odot}$), making them attractive long-term hosts for civilizations given their longevity and stable habitable zones \cite{Loeb2016,Winters2019}. These factors critically influence the temperature and size of a Dyson sphere and thus their observational properties.

Explicit studies \cite[e.g.][]{Loeb2012,Loeb2016} have highlighted the theoretical landscape of Dyson sphere research, emphasizing the suitability of low-mass stars such as red dwarfs for hosting sustainable civilizations and the detection strategies based on combined optical and infrared signatures. These results underscore the importance of correlating megastructure properties with stellar characteristics to improve techno-signature investigations.

The infrared and optical detectability of Dyson spheres surrounding white dwarfs suggests distinct spectral energy distributions that facilitate identification \cite[e.g.][]{Zuckerman2014}. These distributions are enhanced by excess infrared emission, potentially combined with optical observations identifying transit or occultation events by partial megastructures. \cite{Wright2014} described these ideas within the $\hat{G}$ Project framework, outlining systematic methodologies for infrared techno-signature searches and placing constraints on their cosmic prevalence. A modern and comprehensive treatment of Dyson spheres, including full enclosures, partial structures, and their observational implications, is provided by \cite{wright2020}.

In this paper, we investigate a broad parameter space of Dyson sphere radii and their corresponding equilibrium temperatures for host stars ranging from white dwarfs to red M--dwarfs. By calculating the temperature--radius correlation, we place these theoretical megastructures on a Hertzsprung--Russell (H--R) diagram, allowing for direct comparison with normal stellar populations and aiding in the detection of unique thermal signatures. This approach offers new perspectives into the possible observational evidence of Dyson spheres and guides focused strategies for future surveys.

\section{White dwarfs, Red M-dwarfs, and Dyson Spheres}

White dwarf and red M-dwarf stars are particularly intriguing candidates for the construction and detection of Dyson spheres due to their astronomical physical characteristics and their potential for continuous energy acquisition.

Red M-dwarfs, the most abundant stellar type in the Milky Way, comprise approximately 70 percent of all stars \cite{Bochanski2010}. They possess lifespans of trillions of years, far exceeding the present age of the Universe, offering a steady and enduring energy supply for advanced civilizations \cite{Laughlin1997,Loeb2016}. By contrast, white dwarfs are compact remnants of low- and intermediate-mass stars and, despite their low luminosities, radiate steadily for billions of years as they cool \cite{Fontaine2001}. Both white and red M-dwarfs thus demonstrate energetically stable, long-term power supplies for megastructures. We note that this assessment refers to energetic and observational suitability for Dyson sphere construction, and does not imply biological habitability.

From an engineering perspective, the relatively low luminosities of red M-dwarfs ($0.0001$--$0.6\,L_{\odot}$) and white dwarfs ($<0.001\,L_{\odot}$ for most cooling ages) imply that a Dyson sphere could be constructed at a smaller radius than around a Sun-like star while still capturing the stellar energy. For a red M-dwarf, the habitable zone is typically located between 0.05 and 0.3~AU \cite{Kopparapu2013}, allowing a Dyson sphere to be built compactly at moderate material cost. For white dwarfs, the corresponding radius can be a few million kilometers, reducing structural complexity and making full energy capture more feasible \cite{Agol_2011}.

Observationally, these stars are attractive Dyson sphere hosts because their low intrinsic luminosity enhances the contrast between stellar radiation and any waste heat emitted by the structure. A typical Dyson sphere radiating as a blackbody at $\sim$300~K would peak in the mid-infrared (MIR), near 10~$\mu$m. This is within the sensitivity range of missions such as \textit{WISE} and \textit{JWST} \cite{Wright2014}. \cite{Zuckerman2014} have shown that for white dwarfs, any significant infrared excess above the expected cooling curve is particularly suspicious, as few natural processes produce such an effect without accompanying dust signatures. Civilizations aiming for long-term energy sustainability could favorably locate themselves around low-mass, long-lived stars like red M-dwarfs. The smaller size of white dwarfs enables the construction of full energy-capturing spheres at moderate orbital radii, reducing material demands relative to a Sun-like star. These factors make red and white dwarfs attractive hosts for Dyson spheres from an energetic and observational perspective. Detection strategies are enhanced when the host star is faint, increasing the prominence of reprocessed infrared radiation \cite{Wright2014}.

Theoretical placement of Dyson spheres on the H--R diagram provides a diagnostic for anomalous energy distributions. For both white dwarfs and red M-dwarfs, a fully enclosing Dyson sphere effectively blocks stellar luminosity from the diagram, shifting the object to the low-temperature region populated by blackbodies. This displacement is more pronounced for intrinsically faint stars, meaning these stellar types offer the clearest H--R diagram techno-signatures \cite{Timofeev2000}. While equilibrium temperature and luminosity considerations indicate where Dyson spheres may be detectable, additional factors such as surface gravity are required for any habitability assessment. Hence, our study focuses on white dwarfs and red M-dwarfs as crucial host candidates for Dyson spheres.

\section{Dyson Spheres around White and Red M-dwarfs}

The thermal behavior and radiative properties of Dyson spheres have been comprehensively formulated in \cite{wright2020}, which treats both full and partial stellar enclosures. In this work, we adopt the full Dyson sphere limit, in which the structure completely encloses the host star and intercepts its total luminosity. Assuming radiative equilibrium, the stellar luminosity is re-emitted thermally by the Dyson sphere as:

\begin{equation}
L_D = 4 \pi R_D^2 \sigma T_D^4,
\end{equation}

where $R_D$ and $T_D$ are the radius and equilibrium temperature of the Dyson sphere, respectively. Using the stellar luminosity $L_* = 4 \pi R_*^2 \sigma T_*^4$ where $R_*$ and $T_*$ are the radius and  temperature of the star, the equilibrium temperature becomes:

\begin{equation}
T_D = T_* \left(\frac{R_*}{R_D}\right)^{1/2},
\end{equation}

For a full Dyson sphere, the total re-emitted luminosity equals the stellar luminosity, $L_D = L_*$. The observed bolometric flux at a distance $d$ is given by:

\begin{equation}
F(d) = \frac{L_*}{4 \pi d^2},
\end{equation}
appropriate for an unresolved, isotropically emitting source.

Table~\ref{tab:results} summarizes Dyson sphere equilibrium temperatures $T_D$, luminosities $L_D$, and bolometric fluxes $F(d)$ for representative white dwarf and red M-dwarf host stars over a range of Dyson sphere radii. All values now assume a full-sphere energy balance consistent with \cite{wright2020}. As shown, the equilibrium temperature decreases with radius approximately as $T_D \propto R_D^{-1/2}$.

The total luminosity remains fixed by the stellar output, as required for a full Dyson sphere. Larger Dyson spheres emit at longer wavelengths (cooler blackbodies) but with total power determined by the stellar luminosity. Red M-dwarf Dyson spheres have higher luminosities than those around white dwarfs, while white dwarf Dyson spheres produce thermal signatures peaking in the near- to mid-infrared, offering cleaner observational targets.

We emphasize that this analysis considers energetic and observational suitability for Dyson sphere construction, and does not imply biological habitability. Surface gravity and other physical conditions, which are relevant for habitability, are not evaluated in this work.

Figure~2 shows the positions of Dyson spheres on the H--R diagram relative to their host stars. White dwarf and M-dwarf hosts are now explicitly distinguished, and the Dyson sphere locations correspond to the parameter ranges in Table~1. Dyson spheres appear at lower effective temperatures and luminosities due to thermal reprocessing of stellar radiation. This figure, in combination with Table~\ref{tab:results}, provides a clear quantitative framework for evaluating detectability and planning targeted infrared surveys using instruments such as \textit{JWST} and \textit{WISE}.


\begin{figure*}[t]
    \centering
    \includegraphics[width=\linewidth]{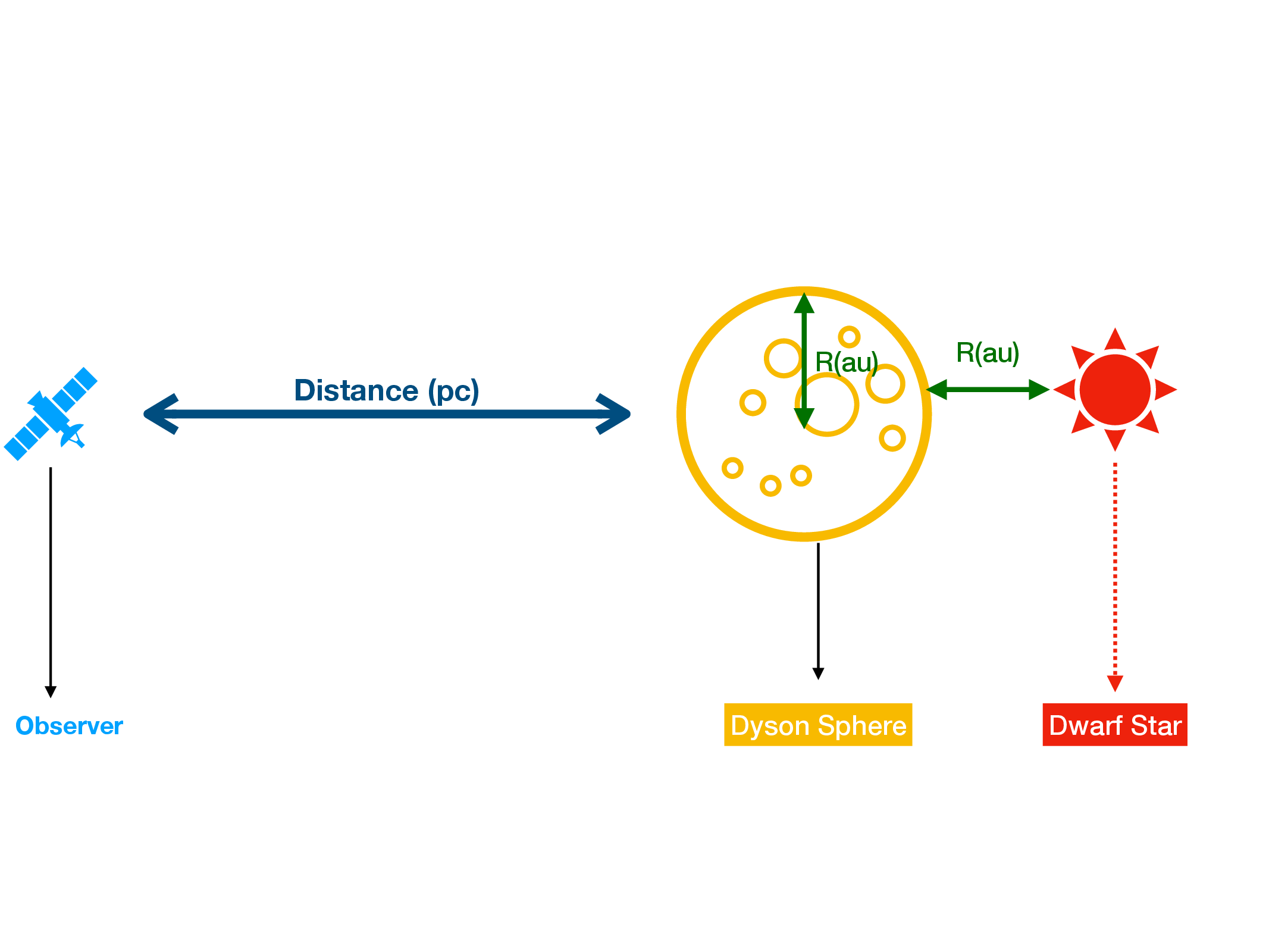}
    \caption{Illustration of a full Dyson sphere around a host star at a given distance (pc) from the observer. The red star represents either a white dwarf or an M-type red dwarf, while the Dyson sphere is depicted in orange to highlight its artificial nature.}
\end{figure*}

\begin{table}[ht]
\centering
\caption{full Dyson sphere temperatures, luminosities, and fluxes at 100 parsec assuming full interception ($f=1$).}
\label{tab:results}
\begin{tabular}{lcccc}
\toprule
Host & $R_{\rm D}$ (AU) & $T_{\rm D}$ (K) & $L_{\rm D}$ (erg s$^{-1}$) & $F(100\,\mathrm{pc})$ (erg s$^{-1}$ cm$^{-2}$) \\
\midrule
\multicolumn{5}{l}{\emph{Red M-dwarf}: $T_* = 3300$ K, $R_* = 1.7 \times 10^{8}$ m} \\
 & 0.5  & 157.32 & $2.44 \times 10^{31}$ & $2.04 \times 10^{-11}$ \\
 & 1.0  & 111.24 & $2.44 \times 10^{31}$ & $2.04 \times 10^{-11}$ \\
 & 5.0  & 49.75  & $2.44 \times 10^{31}$ & $2.04 \times 10^{-11}$ \\
 & 10.0 & 35.18  & $2.44 \times 10^{31}$ & $2.04 \times 10^{-11}$ \\
\midrule
\multicolumn{5}{l}{\emph{White dwarf}: $T_* = 5000$ K, $R_* = 8.4 \times 10^{6}$ m} \\
 & 0.5  & 52.99  & $3.14 \times 10^{29}$ & $2.62 \times 10^{-13}$ \\
 & 1.0  & 37.47  & $3.14 \times 10^{29}$ & $2.62 \times 10^{-13}$ \\
 & 5.0  & 16.76  & $3.14 \times 10^{29}$ & $2.62 \times 10^{-13}$ \\
 & 10.0 & 11.85  & $3.14 \times 10^{29}$ & $2.62 \times 10^{-13}$ \\
\midrule
\multicolumn{5}{l}{\emph{White dwarf}: $T_* = 7000$ K, $R_* = 7.7 \times 10^{6}$ m} \\
 & 0.5  & 71.02  & $1.01 \times 10^{30}$ & $8.47 \times 10^{-13}$ \\
 & 1.0  & 50.22  & $1.01 \times 10^{30}$ & $8.47 \times 10^{-13}$ \\
 & 5.0  & 22.46  & $1.01 \times 10^{30}$ & $8.47 \times 10^{-13}$ \\
 & 10.0 & 15.88  & $1.01 \times 10^{30}$ & $8.47 \times 10^{-13}$ \\
\bottomrule
\end{tabular}
\end{table}

Larger Dyson spheres emit at longer wavelengths (cooler blackbodies) but with total power fixed. Red M-dwarf Dyson spheres have higher luminosities than those around white dwarfs. White dwarf Dyson spheres produce thermal signatures peaking in the near- to mid-infrared and offer cleaner detection targets. These results inform observational strategies for infrared excess searches \cite{Dyson1960,Carrigan2009,Zuckerman2014,Wright2014,Lacki2016}. 

\section{Conclusion}

We have investigated the theoretical properties and observational signatures of the full Dyson sphere around white dwarfs and red M-dwarfs, focusing on equilibrium temperatures, luminosities, and infrared fluxes. A full Dyson sphere would appear at the low-temperature end of the H--R diagram, substantially displaced from their stellar hosts. Equilibrium temperature decreases as $T_{\rm D} \propto R_{\rm D}^{-1/2}$, while total luminosity remains equal to stellar output, leading to emission predominantly in the mid- to far-infrared for the largest structures.

These findings remark the suitability of low-luminosity stars as Dyson sphere hosts, emphasizing that white dwarfs offer clean observational environments due to their simple spectra, which are well described by smooth photospheric or cooling models and contain few intrinsic infrared features, making any excess emission readily identifiable, and that red M-dwarfs provide long-term energetic stability. Predicted fluxes and temperature ranges provide guidance for targeted infrared searches, including with JWST. Future studies combining stellar catalogs, infrared photometry, and synthetic spectral modeling can optimize candidate selection and maximize the efficiency of techno-signature surveys. This work establishes a framework for systematically assessing the detectability of full Dyson sphere.

\section{Future Studies}

Systematic searches for full Dyson spheres can focus on nearby stars from catalogs such as Gaia DR3, TESS Input Catalog (TIC), and infrared surveys like 2MASS and WISE, targeting white dwarfs and red M-dwarfs. Candidates can be examined for infrared excess at $\approx$100--300~K and potential photometric variability if the full sphere occludes the stellar flux. JWST multi-band photometry and low-resolution spectroscopy with MIRI can provide sensitive measurements to detect these signatures and distinguish them from natural sources such as debris disks. Combining catalog information with synthetic spectral modeling allows prioritization of the most promising targets based on distance, luminosity, and stellar temperature, and cross-validation against stars with unusual infrared excess.

Dyson spheres are expected to exhibit smooth, nearly blackbody spectral energy distributions, with peak emission wavelengths determined by their equilibrium temperatures. For temperatures in the range $T_{\rm D} \approx 50$--300~K, the thermal emission peaks between approximately 10 and 60~$\mu$m, placing these objects squarely within the sensitivity range of mid-infrared facilities. Multi-band JWST photometry can identify Dyson sphere candidates through infrared colors inconsistent with stellar photospheres, while low-resolution spectroscopy with MIRI can further discriminate artificial structures from natural sources. In particular, Dyson spheres are expected to lack solid-state spectral features (e.g., silicate emission at 10~$\mu$m) commonly associated with dusty disks, instead showing a smooth continuum consistent with thermal reprocessing of stellar radiation. Such spectral simplicity, combined with an infrared excess and suppressed optical luminosity, provides a distinctive techno-signature diagnostic. For nearby systems ($\lesssim 100$~pc), the predicted infrared fluxes of Dyson spheres around white dwarfs and red M-dwarfs fall within the sensitivity limits of JWST. In particular, JWST/MIRI can detect blackbody emitters at $T_{\rm D} \sim 100$--300~K for distances beyond tens of parsecs, enabling targeted techno-signature searches among nearby low-luminosity stars.

\begin{figure*}[t]
    \centering
    \includegraphics[width=\linewidth]{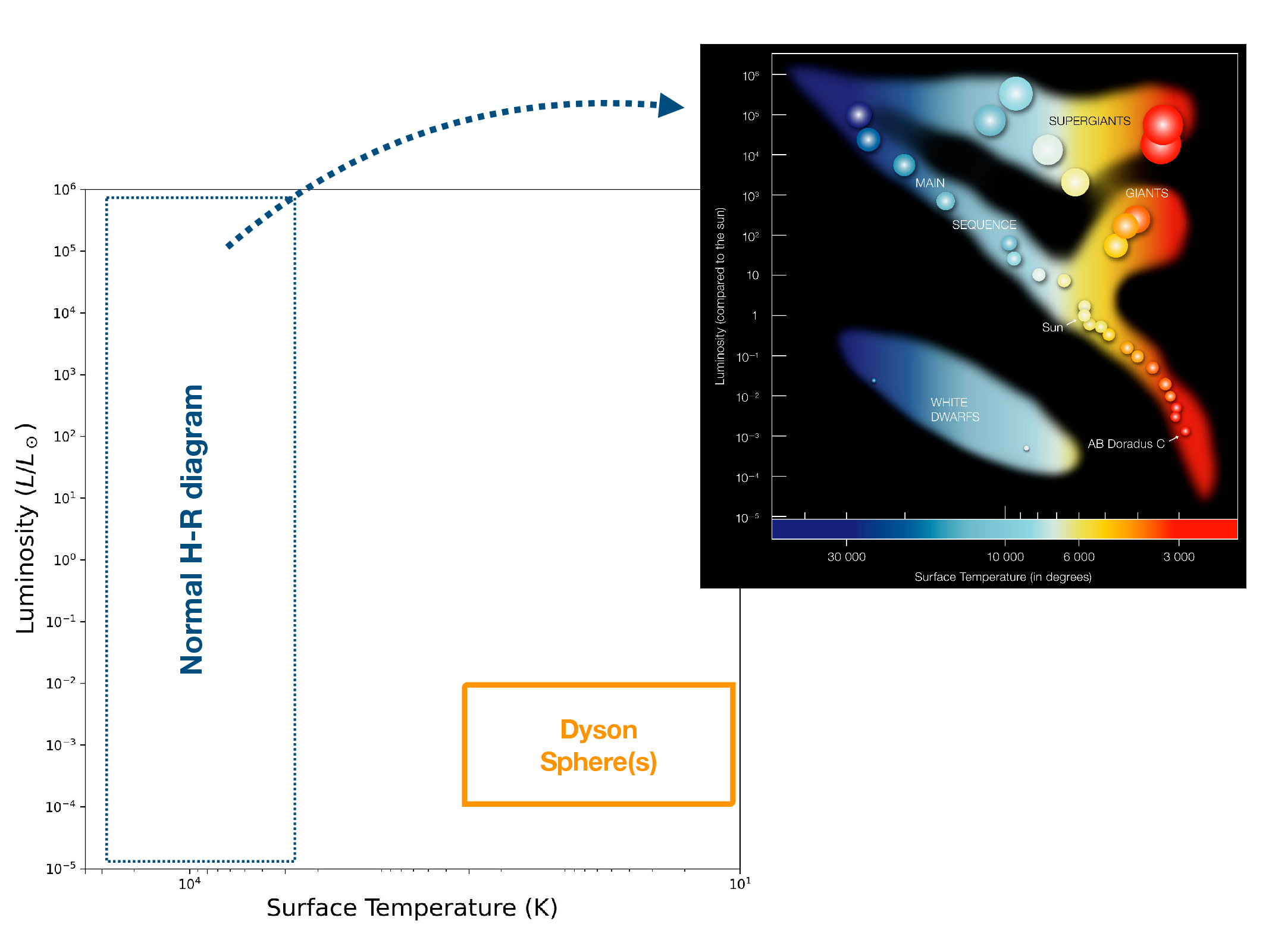}
    \caption{A Schematic H--R diagram showing main-sequence stars, Supergiants, giants, and white dwarfs. Dyson spheres around white dwarfs and red M--dwarfs (orange square area) appear at low temperatures and low luminosities, displaced from their host stars. Top-right (black) figure is reproduced from publicly available materials of the European Southern Observatory (ESO), \texttt{https://www.eso.org/public/images/eso0728c/}.}
\end{figure*}
\section{Acknowledgments}
We express our gratitude to the referees for improving
this paper through their valuable considerations. We are grateful to Avi Loeb (Astronomy Department, Harvard University)for valuable discussions that were critical to this work.
\newpage
\bibliography{main}

\end{document}